\documentclass[twocolumn,epsfig,aps,prb]{revtex4}

\begin{document}

\title{Dephasing of electrons in mesoscopic metal wires due to zero-point fluctuations of optically active localized plasmon modes}

\author{Igor I. Smolyaninov}

\affiliation{Department of Electrical and Computer
Engineering, University of Maryland, College Park, MD 20742, USA}

\date{\today}
\begin{abstract}
Optically active localized plasmon modes may be very abundant on a rough surface of a metal wire. Zero-point fluctuations of these modes are shown to produce local DC magnetization. Thus, the effect of these modes on the electron dephasing time is similar to the effect of magnetic impurities.  
\end{abstract}
\maketitle

Despite years of research, the problem of low temperature behavior of the electron dephasing time $\tau _{\phi}$ in mesoscopic metal wires remains a controversial issue. While theory of electron-electron interactions \cite{1} predicts divergence of $\tau _{\phi}$ at zero temperature, experimental measurements of $\tau _{\phi}$ produce inconclusive results. While some groups report observation of the theoretically predicted $\tau _{\phi}\sim T^{-2/3}$ behavior \cite{2,3}, others observe saturation of the phase coherence time \cite{4,5} at very low temperatures. Various suggestions had been put forward over time in order to explain this apparent controversy (a recent discussion of these suggestions can be found in \cite{3}). One of the most popular explanations of $\tau _{\phi}$ saturation involves very dilute magnetic impurities. Such impurities may cause inelastic spin-flip scattering processes. In addition, scattering of electrons by local magnetic moments destroys the time-reversal symmetry and causes a phase shift between the interfering electron waves. While there exist recent experimental data, which strongly support the magnetic impurity hypothesis \cite{3}, Mohanty and Webb report that their samples were sufficiently clean to reject such an explanation \cite{4,5}. Another potential explanation of $\tau _{\phi}$ behavior, which involves zero-point fluctuations of high energy electromagnetic modes of the metal wire was proposed by Golubev and Zaikin \cite{6}. However, their calculations were criticized in \cite{7}, and comparison of the Golubev and Zaikin predictions with the available experimental data performed by Pierre $et$ $al.$ in \cite{3} seems to indicate disagreement with experiment. This ongoing controversy needs to be resolved since the low-temperature behavior of $\tau _{\phi}$ is of critical importance to mesoscopic physics and its applications.

The goal of this paper is to find a middle ground, which could accommodate all the points of view on low-temperature $\tau _{\phi}$ behavior, and all the experimental results reported so far. We are going to show that zero-point fluctuations of some electromagnetic eigenmodes of a metal wire may induce local DC magnetization, so that the effect of these modes on the electron dephasing time is similar to the effect of dilute magnetic impurities. On the other hand, such electromagnetic eigenmodes of the wire must exhibit optical activity (the left-right asymmetry). Preserving perfect symmetry of the metal wire eliminates such modes, and increases $\tau _{\phi}$. 

As a starting point, we should emphasize that the experimental measurements of $\tau _{\phi}$ are performed on good metals, such as gold, silver, copper, etc. These metals exhibit pronounced surface plasmon resonance, and support large variety of localized and propagating plasmon modes \cite{8,9}. Even though Golubev and Zaikin demonstrated the potential importance of high frequency electromagnetic modes in defining the electron dephasing time, these plasmon modes remain a major missing intrinsic element in most of the current experimental and theoretical studies of the transport and magnetic properties of mesoscopic metallic samples, such as wires, rings, and various other shapes. 
The usual justification for disregarding surface and bulk plasmons in low temperature measurements is that the energy of these modes is large compared to the sample temperature, if the sample size is of the order of a few micrometers. 
However, even if there are no real plasmon quanta in the system, the zero-point fluctuations of the electromagnetic field of all the possible plasmon modes in the sample have to be considered. Localized and propagating surface plasmon (SP) modes seem to be especially important, since the spatial distribution of the electromagnetic field in these modes is very inhomogeneous. The surface plasmon field decays exponentially away from the interface in the case of surface plasmon modes, and away from the localization region in the case of a localized plasmon resonance \cite{8,9}. Electrons can be scattered by the nonzero electromagnetic field in the region of a metal wire, which supports the localized plasmon resonance, even though the number of plasmon quanta in this plasmon mode is zero. 

AC electric field is known to induce DC magnetization in media which exhibit optical activity that is linear in applied magnetic field ($\vec{g}=f\vec{H}$, where $\vec{g}$ is the gyration vector). This effect is usually called the inverse Faraday effect. In an isotropic material this DC magnetization is proportional to the square of the local electric field \cite{10}:   

\begin{equation}
\label{eq1} 
\vec{M}=-\frac{if}{16\pi }\vec{D}\times\vec{D}^{\ast} 
\end{equation}

It is clear from eq.(1) that DC magnetization may only be induced by elliptically polarized electric field. Optically-induced magnetization of non-magnetic media due to inverse Faraday effect has been observed in the experiment \cite{11}. 

Since metals exhibit optical activity which is linear in applied magnetic field \cite{10}: 

\begin{equation} 
\label{eq2} 
f(\omega )= -\frac{4\pi Ne^3}{cm^2\omega ^3}=-\frac{e\omega _p^2}{mc\omega ^3} 
\end{equation}

in the Drude model at $\omega >>eH/mc$ (where $\omega _p$ is the plasma frequency and $m$ is the electron mass) optically induced magnetization of metals should be expected. A localized plasmon mode would induce local DC magnetization of the metal if this mode is elliptically polarized, and thus, exhibit optical activity. Plasmon-induced magnetization has been indeed observed in recent experiments \cite{12}.

A rough surface of a metal wire supports large number of localized plasmon modes. Recent experimental study of local optical activity of rough silver surfaces indicated that very large number of these modes is either left- or right- elliptically polarized, which leads to large local optical activity of the rough metal surface \cite{13}. This plasmon mode chirality is possible because locally the surface roughness features have neither center nor plane of symmetry. However, on the macroscopic scale the surface is not optically active due to statistical averaging of these asymmetries. As a result, we must conclude that even at zero temperature when the number of plasmon quanta is zero in each of these optically active local plasmon modes, zero-point fluctuations of the electric field of these modes produce local DC magnetization. These modes would thus behave as magnetic impurities (In addition, bulk plasmon modes may also exhibit local optical activity in asymmetric experimental configurations, and thus contribute to the concentration of effective magnetic impurities. However, unlike experiments with surface plasmon modes in ref.\cite{13}, no direct experimental observation exists yet on local optical activity of bulk plasmons).

Let us estimate a typical magnetic moment of such an effective magnetic impurity. Let us assume that the defect on the surface of the metal wire has cylindrical shape (Fig.1), and consider the magnetic moment of just one left- of right- circular polarized cylindrical plasmon mode. This magnetic moment of an individual plasmon mode would give us an impression on the order of magnitude of the magnetic moment in the case of an asymmetric defect.

Cylindrical surface plasmons (CSP) which exist on a cylindrical
surface of a cylindrical wire are described by the wave vector $k_{z}$ related
to the CSP propagation in the axial direction of the cylinder, and the
angular quantum number $n$ related to the azimuthal CSP
propagation ($k_{\phi }$) along the cylinder circumference. As a result, the CSP trajectory on a cylindrical surface can be represented as a spiral with the period determined by the CSP
 angular quantum number: for $n=0$ there is no angular momentum and such CSP is analogous to the surface plasmon on a plane surface propagating over the cylinder wall in the axial direction; for
$n>>1$ CSPs strongly rotate around the cylinder and their spectrum
converges to $\omega_{p}/2^{1/2}$ in the large wave vector limit \cite{14}.
The magnetic moment $\mu $ of individual CSP quanta can be
determined as

\begin{equation}
\label{eq3} \mu =-\hbar \left( {\frac{\partial \omega }{\partial
H}} \right)_{H=0}  ,
\end{equation}

where $H$ is the external magnetic field applied in the axial
direction. The estimation of CSP magnetic moment is easy to perform in the two limiting cases $n=0$ and $n\to \infty $ if the nonretarded electrostatic approximation is used. These states correspond to the SPP with large wave
vectors, which propagate parallel and perpendicular to the axial
magnetic field, respectively. In this case we can neglect the
effects of curvature of the cylinder, and using the results of \cite{15}, obtain the following expression for the eigenfrequencies of the CSPs:

\begin{equation}
\label{eq4} \omega = \left\{
\begin{array}{l}
(\omega _p ^2+\omega _c
^2)^{1/2}/2^{1/2} \quad \quad \quad \quad {\text for} \quad n=0,\\
(\omega _p ^2+\omega _c ^2)^{1/2}/2^{1/2}\pm \omega _c /2 \quad {\text  for } \quad n\ne 0 . \\
\end{array} \right.
\end{equation}

It is evident that in accordance with eq.(1), the $n=0$ plasmons do not have magnetic moments, while circularly polarized $n\ne 0$ CSP quanta have magnetic moments of the order of the Bohr's magneton $\mu _B$. Exact values of the CSP magnetic moments can be determined by taking into account the effects of curvature of the cylinder, retardation, and the Aharonov-Bohm effect on the CSP dispersion. Analysis of the numerical calculations of the CSP dispersion for a number of different geometries reported in \cite{16,17} indicates that the results of the simple qualitative estimate above remain valid in all these cases: CSPs with zero angular momentum have no magnetic moment, while $n\ne 0$
CSPs have $\mu \sim \mu _{B}$. Magnetization of the same order of magnitude can be expected in the ground state of the surface plasmon field. Thus, optically active sites on the metal surface exhibit magnetization of the same order of magnitude as regular magnetic impurities, and their effect on the electron dephasing time $\tau _{\phi}$ should be similar. 

The concentration of the effective magnetic impurities depends on the surface roughness of the metal wires. Unfortunately, this experimental parameter was not controlled in the decoherence time measurements reported so far. It is impossible to estimate the range of effective concentrations on the basis of available experimental data. On the other hand, the data in Figs. 3 and 4 from ref.\cite{13} indicate that this concentration may reach values of at least a few tens of effective magnetic impurities per square micrometer on a rough surface of silver. In principle, effective impurities concentration may be even higher, since the localization length of localized surface plasmon modes may be as small as a few nanometers \cite{18}. A straightforward experiment performed with good surface roughness control (for example, STM maps of surface topography of the samples may be measured) would give a clear indication of the possible role of plasmon-induced magnetization in mesoscopic experiments.

Another important question regarding electron interaction with these effective magnetic impurities is whether the plasmon-induced magnetization can rotate (or flip) in a scattering process with an electron. This question is important because while contribution of weak localization to the magnetoresistance can be suppressed by both spin-flip scattering and by static, randomly oriented, frozen magnetic moments, the universal conductance fluctuations are not suppressed by the frozen magnetic impurities.  In ref.\cite{5} Mohanty and Webb observe that the phase coherence time deduced from the universal conductance fluctuation measurements in large magnetic field also saturates at low temperature. In this work they assumed that the magnetic moments of the impurities in the sample are frozen at low temperature and applied large magnetic field and, thus, become static and should not contribute to the universal conductance fluctuations. While the electromagnetic energy of the zero-point fluctuations could not change in a scattering process with an electron (it always remains equal to $1/2\hbar \omega $), the phase of the zero-point electric field may change due to scattering. If the linewidth of the localized SP mode would be zero, such change in phase would not lead to a different value of $\vec{M}$ resulting from eq.(1). In reality, the linewidth of every SP mode is always nonzero, and the DC magnetization defined by eq.1 always fluctuates (in fact, the term DC in this case means only that the characteristic frequency of these fluctuations is much smaller then $\omega $). As a result, the phase change of the zero-point electric field due to scattering would lead to a different value of the local magnetization, which is seen by the electron moving along the time-reversed trajectory, and thus, to the dephasing effect. This dephasing mechanism cannot be frozen by an external magnetic field in accordance with the observation in \cite{5}. The effective magnetic impurities due to the chiral plasmon modes will always behave like regular magnetic impurities above the Kondo temperature.

In summary, we have considered the effect of optically active localized plasmon modes on the electron dephasing time in mesoscopic wires. Such modes may be very abundant on a rough surface of a metal wire. Zero-point fluctuations of these modes are shown to produce local DC magnetization. The effect of these modes on the electron dephasing time is similar to the effect of magnetic impurities.  
Thus, improvement of the electron dephasing time in a mesoscopic metal wire would require not only elimination of regular magnetic impurities, but improvement of the shape of the wire and its surface as well.

This work has been supported in part by the NSF grant ECS-0304046.

\begin{figure}
\caption{Model geometry for the calculations of the magnetic moments of localized surface plasmon modes of a surface defect. Note that in order to exhibit local optical activity, and possess a total magnetic moment, the defect must be asymmetric, which is not the case in this model geometry.} 
\label{fig1}
\end{figure}

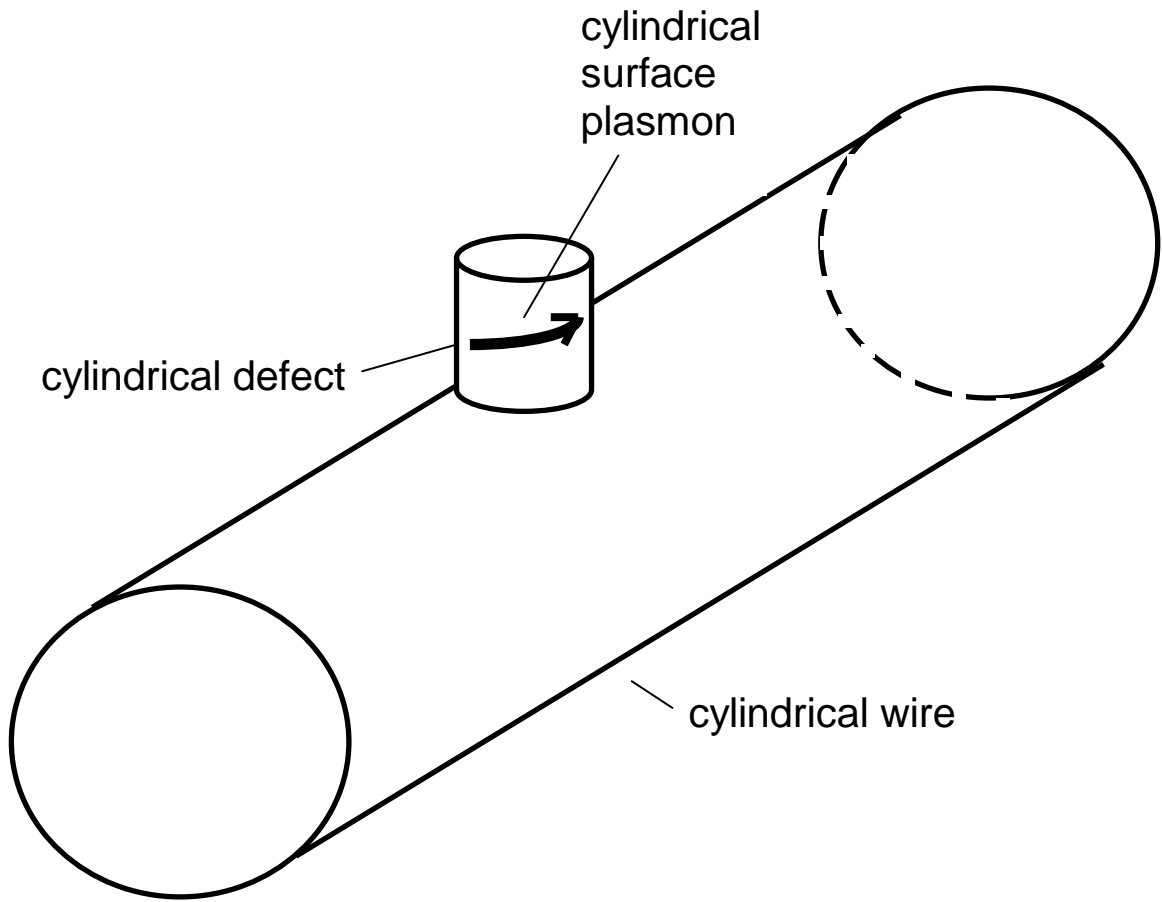
\end{document}